\documentclass{nature}
\usepackage{epsfig}
\usepackage{amsmath}
\usepackage{amsmath,mathrsfs,amsbsy,color,graphicx,bm,amsthm,amsfonts}
\usepackage{units}
\usepackage{bbm}
\usepackage{times}
\usepackage{dcolumn}
\usepackage{mathrsfs}
\usepackage{amsmath,amssymb,epsfig,float}

\begin{document}

\title{Electric-field-induced interferometric resonance of a one-dimensional
spin-orbit-coupled electron}
\author{Jingtao Fan$^{1,2}$, Yuansen Chen$^{3,2}$, Gang Chen$^{1,2,*}$,
Liantuan Xiao$^{1,2}$, Suotang Jia$^{1,2}$, Franco Nori$^{4,5}$ }
\maketitle

\begin{affiliations}
\item
State Key Laboratory of Quantum Optics and Quantum Optics Devices, Institute of Laser spectroscopy, Shanxi University, Taiyuan 030006, China \\
\item
Collaborative Innovation Center of Extreme Optics, Shanxi University,
Taiyuan, Shanxi 030006, China \\
\item
State Key Laboratory of Quantum Optics and Quantum Optics Devices, Institute
of Opt-Electronics, Shanxi University, Taiyuan, Shanxi 030006, China\\
\item
CEMS, RIKEN, Saitama 351-0198, Japan\\
\item
Department of Physics, University of Michigan, Ann Arbor, Michigan
48109-1040, USA\\ \newline
$^*$Corresponding authors, e-mail: chengang971@163.com\newline
\end{affiliations}

\begin{abstract}
The efficient control of electron spins is of crucial importance for
spintronics, quantum metrology, and quantum information processing. We
theoretically formulate an electric mechanism to probe the electron spin
dynamics, by focusing on a one-dimensional spin-orbit-coupled nanowire
quantum dot. Owing to the existence of spin-orbit coupling and a pulsed
electric field, different spin-orbit states are shown to interfere with each
other, generating intriguing interference-resonant patterns. We also reveal
that an in-plane magnetic field does not affect the interval of any
neighboring resonant peaks, but contributes a weak shift of each peak, which
is sensitive to the direction of the magnetic field. We find that this
proposed external-field-controlled scheme should be regarded as a new type
of quantum-dot-based interferometry. This interferometry has potential
applications in precise measurements of relevant experimental parameters,
such as the Rashba and Dresselhaus spin-orbit-coupling strengths, as well as
the Land\'{e} factor.
\end{abstract}

\baselineskip=0.8 cm

Being an intrinsic property of condensed-matter materials, spin-orbit
coupling (SOC) mixes the orbital and spin degrees of particles, and opens
the possibility of electric control of the electron spin via its orbit,
apart from the well-known magnetic responses\cite%
{KFY07,YXE12,SYH13,SB10,JPA14,RL62,EA03,VMD06,LRYJ13,MT08,TJA14,JDE16,JGA15,MTY08,SV12,NFB10,VSS13}%
. A notable example exploiting SOC in semiconductor nanostructures is called
the electric-dipole spin resonance technique\cite{RL62,EA03,VMD06,LRYJ13}
(EDSR), in which a spin-orbit qubit is encoded into a SOC-hybridized spin
doublet and an oscillating electric field is further applied to manipulate
this qubit on its Bloch sphere. Recently, much theoretical\cite%
{LRYJ13,JDE16,JGA15} and experimental\cite{MTY08,SV12,NFB10,VSS13} attention
have been paid to explore the EDSR in semiconductor quantum dot (QD). For
example, utilizing this technique, the single spin-orbit qubit operation has
been achieved\cite{SV12} and the spin-orbit effective field can also be
determined\cite{NFB10}, which reflects its potential application in quantum
information processing and parameters measurement. In addition, the
SOC-assisted spin control, such as the magnetic-free spin filtering\cite%
{SB10,JPA14} where the SOC serves as a necessary ingredient to spatially and
electrically separate electrons with different spins, has also been
achieved. In contrast to the conventional fully-magnetic control, the
introduction of electric passage via SOC paves a much more experimentally
feasible way to locally address electron spin, which may impact spintronics%
\cite{Spintronics}.

Matter-wave interference exquisitely exhibits the wave nature of particles,
which offers microscopic information of certain physical processes\cite%
{SPS12,AJD09,RLF06,JYJ12,SN12}. Various interferometries, for, e.g. electrons%
\cite{ML53,ML54,BJ09,YYD03,ERM97,DGK08,LA10,SHH09,JS12}, neutrons\cite%
{RWU74,RH00}, and atoms\cite{PRB97,JJ99,MA15}, have been widely applied to
measure various physical quantities, by virtue of their wave nature. With
its rapid improvement of relevant experiment and theory, SOC, which exists
naturally in condensed-matter systems\cite{Rashba04,Wolf01,Nitta97} and is
also simulated in ultracold atomic systems\cite{HuiZ,ISpielman,JD11}, is
expected to be a new physical resource to demonstrate particle coherence in
a spin-orbit-mixed way.

In this report, we theoretically formulate an electric mechanism to
interfere electron orbits, by focusing on a one-dimensional (1D)
spin-orbit-coupled nanowire QD. Owing to the existence of SOC and a pulsed
electric field, different spin-orbit states are shown to interfere with each
other, generating intriguing interference-resonant patterns. Furthermore, an
in-plane magnetic field, treated as a perturbation, is also introduced to
probe the relevant dynamics. We find that this magnetic field does not
affect the interval of any neighboring resonant peaks, but contributes a
weak shift of each peak, which is sensitive to the direction of the applied
magnetic field. We find that this proposed external-field-controlled scheme,
exhibiting all the basic ingredients of a quantum interferometer, should be
regarded as multi-arm interferometry. We emphasize that the obtained
interferometric signal originates from the out-of-phase interference of the
dynamical phase factors of the infinite spin-orbit states, which is
remarkably different from conventional optical/atomic interferometers\cite%
{JJ99}. This interferometry has potential applications in precise
measurements of relevant experimental parameters, such as the Rashba and
Dresselhaus SOC strengths, as well as the Land\'{e} factor.\newline

\section*{{\protect\LARGE \textbf{Results}}}

The system we consider is a 1D nanowire QD with SOC, confined in a harmonic
well and subjected to time-dependent external electric and magnetic fields.
The total Hamiltonian can be divided into three parts\cite{LRYJ13,TJA14}%
\begin{equation}
H=H_{\text{0}}+H_{\text{E}}+H_{\text{Z}}.  \label{HT}
\end{equation}%
Here, the \textquotedblleft free\textquotedblright\ Hamiltonian, without
external fields, reads%
\begin{equation}
H_{\text{0}}=\frac{p^{2}}{2m}+\frac{1}{2}m\omega ^{2}x^{2}+\alpha _{R}\sigma
_{y}p+\alpha _{D}\sigma _{x}p,  \label{H0}
\end{equation}%
where $p=-i\hbar \partial /\partial x$, $m$ is the effective electron mass, $%
\alpha _{R(D)}$ is the Rashba (Dresselhaus) SOC strength, and $\sigma
_{x(y)} $ is the Pauli spin operator. The Hamiltonian for the
electric-dipole energy, induced by an external electric field $\mathcal{E}%
(t) $, is written as
\begin{equation}
H_{\text{E}}=e\mathcal{E}(t)x.  \label{HE}
\end{equation}%
The Hamiltonian for the Zeeman energy of an electron, under an in-plane
magnetic field $\mathbf{B}(t)$, is given by
\begin{equation}
H_{\text{Z}}=\frac{1}{2}g_{e}\mu _{B}\mathbf{B}(t)\cdot \mathbf{\sigma ,}
\label{HZ}
\end{equation}%
where $g_{e}$ is the Land\'{e} factor, $\mu _{B}$ is the Bohr magneton, and $%
\mathbf{B}(t)=B(t)\mathbf{n}$, with $\mathbf{n=(}\cos \theta \mathbf{,}\sin
\theta ,0\mathbf{)}$ being the direction of the external magnetic field.

It is convenient to introduce two auxiliary parameters,
\begin{equation}
\alpha =\sqrt{\alpha _{R}^{2}+\alpha _{D}^{2}}\text{, \ \ \ }\varphi
=\arctan (\frac{\alpha _{R}}{\alpha _{D}})\text{,}  \label{AP}
\end{equation}%
to rotate the spin space, along the $z$ axis, to a new frame. In this case,
the Hamiltonians (\ref{H0}) and (\ref{HZ}) become
\begin{equation}
H_{\text{0}}=\frac{p^{2}}{2m}+\frac{1}{2}m\omega ^{2}x^{2}+\alpha \Sigma
_{x}p,\;  \label{TH0}
\end{equation}%
and%
\begin{equation}
H_{\text{Z}}=\frac{1}{2}g_{e}\mu _{B}B(t)\left[ \Sigma _{x}\cos (\theta
-\varphi )+\Sigma _{y}\sin (\theta -\varphi )\right] ,  \label{THz}
\end{equation}%
where $\Sigma _{x}=\sigma _{x}\cos \varphi +\sigma _{y}\sin \varphi $ and $%
\Sigma _{y}=-\sigma _{x}\sin \varphi +\sigma _{y}\cos \varphi $ are the
redefined spin operators in the new frame, while the Hamiltonian (\ref{HE})
remains unchanged. Since SOC endows the quantum dot with the ability to
respond to both the external electric and magnetic fields, our goal here is
to build a new-type of quantum-dot-based interferometer by utilizing this
natural response.

Before specifying the temporal shapes of the external fields, we first
analyze the \textquotedblleft unpertubed\textquotedblright\ Hamiltonian $%
H_{0}$, under which the initial state is prepared. Taking into account the
conservation of the redefined spin operator $\Sigma _{x}$, the eigenstates
of the Hamiltonian $H_{\text{0}}$ are represented as
\begin{equation}
\left\vert \phi _{n}^{\sigma }\right\rangle \left\vert \sigma \right\rangle
=\exp \left( -\frac{i}{\hbar }m\alpha \Sigma _{x}\cdot x\right) \left\vert
\phi _{n}\right\rangle \left\vert \sigma \right\rangle ,  \label{WF}
\end{equation}%
where the orbital part $\left\vert \phi _{n}\right\rangle $ is the $n$th
eigenstate of a harmonic oscillator and the spin part $\left\vert \sigma
\right\rangle $ is the eigenstate of the redefined spin operator $\Sigma
_{x} $, i.e., $\Sigma _{x}\left\vert \sigma \right\rangle =\sigma \left\vert
\sigma \right\rangle $, with $\sigma =\pm 1$. Notice that for each electron
orbit, the total eigenstates are twofold degenerate. We assume that the
\textquotedblleft unpertubed\textquotedblright\ system is initially prepared
in its ground state ($n=0$), with a general superposition of two spin
components, i.e., $\left\vert \Psi (0)\right\rangle =c_{+}\left\vert \phi
_{0}^{+}\right\rangle \left\vert +\right\rangle +c_{-}\left\vert \phi
_{0}^{-}\right\rangle \left\vert -\right\rangle $, where $\left\vert
c_{+}\right\vert ^{2}+\left\vert c_{-}\right\vert ^{2}=1$.

To run the dynamics, we turn-on the external fields at a certain time $t_{0}$%
. In our proposed interferometer, the external fields are utilized as
\textquotedblleft phase objects\textquotedblright\ to generate proper
interferometric phases\cite{JJ99}, and their detailed field profiles should
be well engineered. As an instructive example, the electric and magnetic
signals are now taken as
\begin{equation}
B(t)=B_{0}\Theta (t-t_{0}),  \label{BT}
\end{equation}%
\begin{equation}
\mathcal{E}(t)=\mathcal{E}_{0}\exp \left[ -\frac{(t-t_{0})^{2}}{\sigma
_{t}^{2}}\right] ,  \label{ET}
\end{equation}%
where $\Theta (t-t_{0})$ is the Heaviside step function, and $\mathcal{E}%
_{0} $ and $\sigma _{t}$ are the peak amplitude and temporal width of the
Gaussian-type pulse, respectively. The above expressions of the external
fields show that the magnetic field, characterized by the constant field
strength $B_{0}$, looks like a simple quantum \textquotedblleft quenching
knob\textquotedblright , whose effect is quite different from the electric
field. This different choice of the electric and magnetic fields relies on
their individual roles in activating novel dynamics in the nanowire QD, as
will be described below.

To further facilitate the theoretical description, we prefer to transform
the Hamiltonian (\ref{HT}) to the frame of the \textquotedblleft
velocity\textquotedblright\ gauge by using a unitary operator $U=\exp \left[
\frac{i}{\hbar }eA(t)\cdot x\right] $, with the gauge potential%
\begin{equation}
A(t)=\int_{0}^{t}\mathcal{E}(\tau )d\tau =\frac{1}{2}A_{0}\left[ \text{erf}%
\left( \frac{t-t_{0}}{\sigma _{t}}\right) +\text{erf}\left( \frac{t_{0}}{%
\sigma _{t}}\right) \right] ,  \label{GP}
\end{equation}%
where $A_{0}=\sqrt{\pi }\sigma _{t}\mathcal{E}_{0}$ and erf$(x)$ is the
error function. After performing the transformation $H\rightarrow UHU^{\dag
}+i\hbar \dot{U}U^{\dag }$, the Hamiltonian (\ref{HT}) becomes%
\begin{equation}
H=\frac{p^{2}}{2m}+\frac{1}{2}m\omega ^{2}x^{2}+\alpha \Sigma _{x}p-\frac{e}{%
m}A(t)p+\frac{1}{2}g_{e}\mu _{B}B(t)\left[ \Sigma _{x}\cos (\theta -\varphi
)+\Sigma _{y}\sin (\theta -\varphi )\right] .  \label{HTN}
\end{equation}%
Despite of its simple form, the Hamiltonian (\ref{HTN}) still governs a
quite complicated dynamics; so much so that no exact solution can be found.
To catch the basic idea of the proposed interferometric process, we simplify
the analysis over two aspects. Firstly, we restrict the Zeeman energy, $%
\delta _{Z}=\frac{1}{2}g_{e}\mu _{B}B_{0}$, to a weak regime such that it is
much less than the orbital splitting, i.e., $\delta _{Z}\ll \hbar \omega $.
Therefore, the Hamiltonian contributed by the magnetic field can be treated
perturbatively. Secondly, from the expression of $A(t)$, we find that in the
limit $\sigma _{t}\rightarrow 0$ and $\mathcal{E}_{0}\rightarrow \infty $ ($%
A_{0}=\sqrt{\pi }\sigma _{t}\mathcal{E}_{0}$ still remains finite), the
gauge potential tends to be $A(t)=A_{0}\Theta (t-t_{0})$, and the
corresponding electric field becomes a delta-type pulse, i.e., $\mathcal{E}%
(t)=A_{0}\delta (t-t_{0})$. It follows that under such a condition, both the
electric and magnetic manipulations become external quantum quenching knobs,
and furthermore, this ultrafast limit of the electric-field pulse allows us
to obtain a compact analytical solution, which captures the key aspects of
this interferometer.

Since the control parameters $A(t)$ and $B(t)$ are switched on at time $%
t_{0} $ and thereafter remain constant, the total dynamics can be
conveniently simplified to two stationary problems of times $t<t_{0}$ and $%
t>t_{0}$, which are governed respectively by the Hamiltonians $H(t<t_{0})$
and $H(t>t_{0})$. We now employ perturbation theory to solve these. We first
concentrate on the zeroth-order Hamiltonian $H_{\text{S}}=H_{\text{0}}+H_{%
\text{E}}$. Notice that without the perturbed magnetic field, the redefined
spin operator $\Sigma _{x}$ commutes with the Hamiltonian $H_{\text{S}}$. In
terms of this, the eigenstates of the Hamiltonian $H_{\text{S}}(t>t_{0})$
are obtained exactly by
\begin{equation}
\left\vert \psi _{n}^{\sigma }\right\rangle \left\vert \sigma \right\rangle
=\exp \left[ -\frac{i}{\hbar }(m\alpha \Sigma _{x}-eA_{0})\cdot x\right]
\left\vert \phi _{n}\right\rangle \left\vert \sigma \right\rangle ,
\label{TWF}
\end{equation}%
with eigenenergies%
\begin{equation}
\epsilon _{n,\sigma }=n\hbar \omega +\sigma e\alpha A_{0}-\frac{1}{2}m\alpha
^{2}-\frac{1}{2}\frac{e^{2}A_{0}^{2}}{m}.  \label{EN}
\end{equation}%
Obviously, the electric field lifts the degeneracy of each orbital energy by
$2e\alpha A_{0}$. However, there exists a special case where the $n$th and ($%
n+k$)th orbits are degenerate or quasi-degenerate, say $\left\vert \epsilon
_{n,+}-\epsilon _{n+k,-}\right\vert \ll \hbar \omega $\ (see the Methods
Section). In such case, the non-degenerate perturbative formula breaks down
and we must use a degenerate perturbation method. Therefore, the complete
solutions should be divided into a nondegenerate case (NC) and a
degenerate/quasi-degenerate case (DC). After a straightforward calculation,
the eigenstates of the Hamiltonian $H(t>t_{0})$, which are accurate up to
first order in the Zeeman energy $\delta _{Z}$, can be summarized as
\begin{equation}
\left\vert \psi _{n,\sigma }\right\rangle =\left\vert \psi _{n}^{\sigma
}\right\rangle \left\vert \sigma \right\rangle +\delta _{Z}\sin (\theta
-\varphi )\overset{\infty }{\sum_{l=0}}\frac{\left\langle \psi _{l}^{-\sigma
}\right\vert \left. \psi _{n}^{\sigma }\right\rangle }{(n-l)\hbar \omega
-2\sigma m\alpha A_{0}}\left\vert \psi _{l}^{-\sigma }\right\rangle
\left\vert \sigma \right\rangle
\end{equation}%
\ for the NC and%
\begin{equation}
\left\vert \psi _{n,\sigma }\right\rangle =a_{n,\sigma }\left\vert \psi
_{n}^{+}\right\rangle \left\vert +\right\rangle +\overset{\infty }{%
\sum_{l\neq n}}F_{1}^{\sigma }(l,n)\left\vert \psi _{l}^{+}\right\rangle
\left\vert +\right\rangle +b_{n+k,\sigma }\left\vert \psi
_{n+k}^{-}\right\rangle \left\vert -\right\rangle +\overset{\infty }{%
\sum_{l\neq n}}F_{2}^{\sigma }(l,n)\left\vert \psi _{l}^{-}\right\rangle
\left\vert -\right\rangle
\end{equation}%
\ for the DC, where $a_{n,\sigma }$, $b_{n,\sigma }$,\ and $F_{i}^{\sigma
}(l,n)$\ ($i=1,2$) are given in the Methods Section. Furthermore, the
corresponding perturbative eigenenergies are given by (see also the Methods
Section for details)%
\begin{equation}
E_{n,\sigma }=\left\{
\begin{array}{c}
n\hbar \omega +\sigma e\alpha A_{0}+\sigma \delta _{Z}\cos (\theta -\varphi )%
\text{ \ (NC)} \\
\\
n\hbar \omega +\frac{1}{2}\sigma (k\hbar \omega -f)\text{ \ \ \ \ \ \ \ \ \
\ \ (DC)}%
\end{array}%
\right. ,  \label{SPEC}
\end{equation}%
where%
\begin{equation}
f=\left\{ 4\delta _{Z}^{2}\left\vert \eta \right\vert ^{2}\sin ^{2}(\theta
-\varphi )+\left[ k\hbar \omega +2e\alpha A_{0}+2\delta _{Z}\cos (\theta
-\varphi )\right] ^{2}\right\} ^{\frac{1}{2}},
\end{equation}%
with $\eta =\left\langle \psi _{n+k}^{-}\right\vert \left. \psi
_{n}^{+}\right\rangle $. Notice that, for simplicity, here we have neglected
the dependence of $f$ on $n$, since $\eta $ is very small for any $n$ and $k$%
. Equation (\ref{SPEC}) shows that the external electric and magnetic fields
dominate the dynamics through different ways: the former (electric) appears
as a weight factor of the SOC strength, whereas the effect of the magnetic
field depends crucially on its specific direction. These features clearly
signal their quite different roles in controlling the interference pattern.

\begin{figure}[t]
\centering\includegraphics[width = 6.3in]{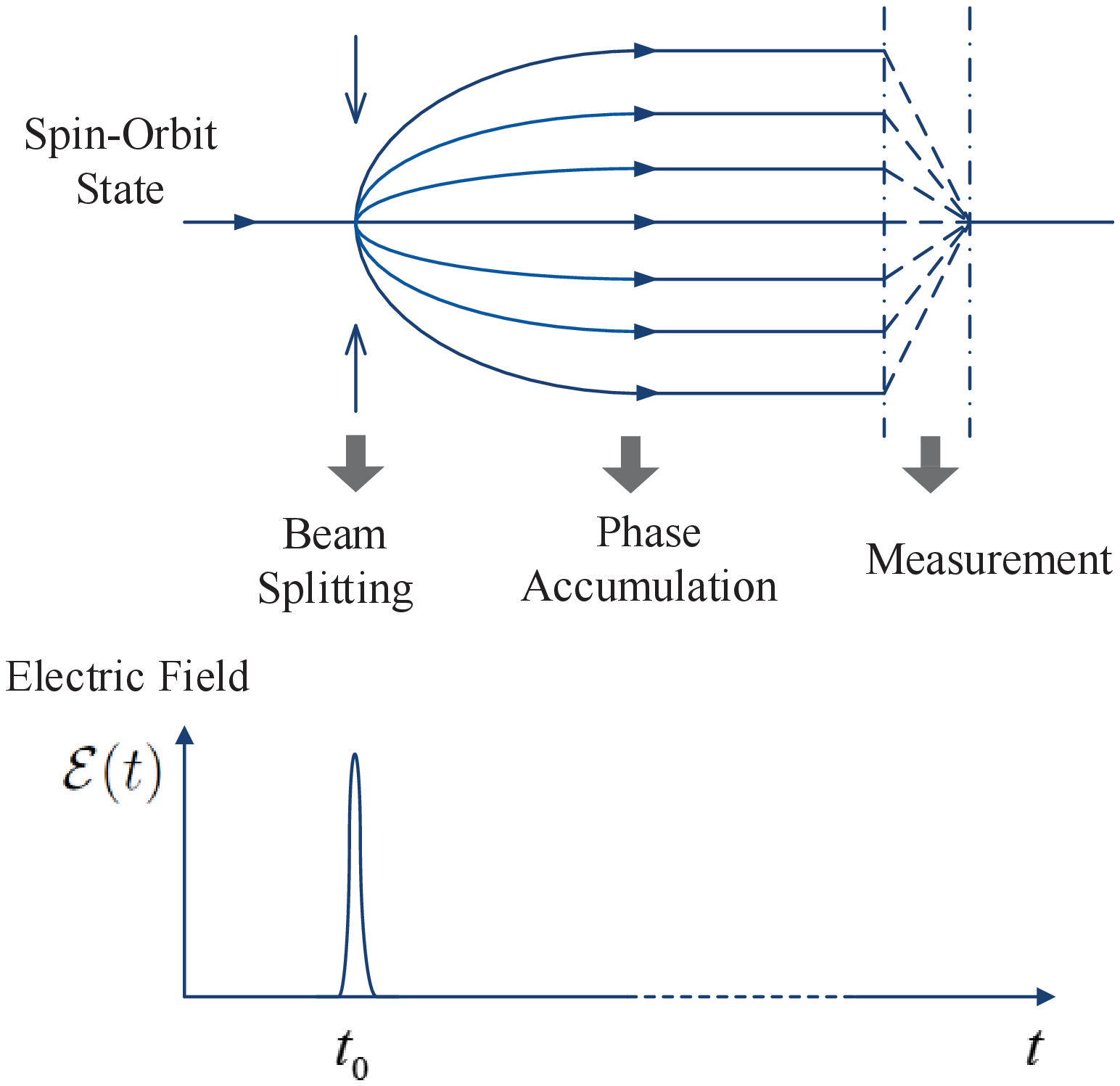}
\caption{\textbf{Schematic description of the proposed interferometer.}}
\label{Draft}
\end{figure}

Having obtained the complete eigenstates and eigenenergies of the quenched
Hamiltonian $H(t>t_{0})$, we are now able to discuss the total dynamics of
the system. After the quantum quench, the whole information of the nanowire
QD with SOC is encoded in its instantaneous wavefunction, which can be
expanded using the spin-orbit basis $\left\vert \psi _{n,\sigma
}\right\rangle $, i.e.,%
\begin{equation}
\left\vert \Psi (t)\right\rangle =\overset{\infty }{\sum_{N=0}}\left[
\overline{A}_{N}\left\vert \psi _{N,+}\right\rangle \exp \left( -\frac{i}{%
\hbar }E_{N,+}t\right) +\overline{B}_{N}\left\vert \psi _{N,-}\right\rangle
\exp \left( -\frac{i}{\hbar }E_{N,-}t\right) \right] ,
\end{equation}%
where $\overline{A}_{N}=\left\langle \psi _{N,+}\right\vert \left. \Psi
(0)\right\rangle $ and $\overline{B}_{N}=\left\langle \psi _{N,-}\right\vert
\left. \Psi (0)\right\rangle $ are the projected coefficients for the
\textquotedblleft $+$\textquotedblright\ and \textquotedblleft $-$%
\textquotedblright\ spin sectors of the $N$th orbit, respectively. A crucial
point we should notice is that being a direct consequence of the pulsed
electric field, $\overline{A}_{N}$ and $\overline{B}_{N}$ may acquire
non-zero values even for $N\neq 0$. In other words, it is the pulsed
electric field at time $t_{0}$ that splits the original zeroth-orbit
wavefunction into other different orbital states $\left\vert \psi _{n,\sigma
}\right\rangle $. Indeed, in the nanowire QD with SOC, we have built a
\textit{multiple-polarization-interferometer}, where the interferometric
arms correspond to infinite different orbital states, and the beam splitter
corresponds to the pulsed electric field (see Fig.~\ref{Draft} for a more
intuitive description). However, to obtain a signature of interference,
usually observed as a population difference of a physical quantity, a
special operation to recombine the split orbital states is still needed.
Motivated by the fact that any different spin-reversed orbital states are
non-orthogonal ($\left\langle \psi _{n}^{-\sigma }\right\vert \left. \psi
_{n^{\prime }}^{\sigma }\right\rangle \neq 0$) due to the existence of SOC,
it is thus convenient to investigate the ensemble average of the spin
polarization $\sigma _{z}$. The result is given by
\begin{eqnarray}
\left\langle \sigma _{z}(t)\right\rangle &=&\overset{\infty }{\sum_{N=0}}%
\overset{\infty }{\sum_{L=0}}\left\{ M_{+-}(N,L)\exp \left[ -\frac{i}{\hbar }%
(E_{N,+}-E_{L,-})t\right] \right.  \label{SZZ1} \\
&&+M_{-+}(N,L)\exp \left[ -\frac{i}{\hbar }(E_{N,-}-E_{L,+})t\right]  \notag
\\
&&+M_{--}(N,L)\exp \left[ -\frac{i}{\hbar }(E_{N,-}-E_{L,-})t\right]  \notag
\\
&&\left. +M_{++}(N,L)\exp \left[ -\frac{i}{\hbar }(E_{N,+}-E_{L,+})t\right] +%
\text{c.c.}\right\} ,  \notag
\end{eqnarray}%
\newline
where $M_{\alpha \beta }(N,L)$ ($\alpha ,\beta =\pm $) are given in the
Methods Section and c.c.~denotes the complex conjugate.

As shown in Eq.~(\ref{SZZ1}), $\left\langle \sigma _{z}(t)\right\rangle $
exhibits the inner product between different spin-reversed orbital states,
namely $\left\langle \psi _{n}^{-\sigma }\right\vert \left. \psi _{n^{\prime
}}^{\sigma }\right\rangle $, which is of great importance to support the
accumulated dynamical phase factors. Evidently, these phase factors will
lead to periodic oscillations of $\left\langle \sigma _{z}(t)\right\rangle $
over time, which may be referred to as (time-domain) interference fringes.
However, instead of focusing on the time-dependent interference signals,
usually done in time-domain Ramsey-like atom interferometry\cite%
{RAMSEY,RAMSEY2}, we prefer to explore the long-time average of $%
\left\langle \sigma _{z}(t)\right\rangle $,
\begin{equation}
Q=\frac{1}{T}\int_{t_{0}}^{t_{0}+T}\left\langle \sigma _{z}(t)\right\rangle
dt,  \label{QT}
\end{equation}%
where $T$ is a long timespan, to extract a more prominent
interference-resonant effect.

Figure 2 shows the evolution of $Q$ versus $E_{0}$, by numerical
integration of the time-dependent Schr\"{o}dinger equation, with $\left\vert
\Psi (0)\right\rangle =$ $\frac{1}{\sqrt{2}}(\left\vert \phi
_{0}^{+}\right\rangle \left\vert +\right\rangle +\left\vert \phi
_{0}^{-}\right\rangle \left\vert -\right\rangle )$ and $T=40\pi /\omega $
(blue-solid curve). It is remarkable to see that the interference pattern
appears, and more interestingly, some sharp interference-resonant peaks are
formed periodically at
\begin{equation}
\frac{\mathcal{E}_{0}}{\hbar \omega /2e\sqrt{\pi \hbar /m\omega }}=n\cdot
250=0,250,500,750,\cdots .
\end{equation}%
The underlying physics of such resonant effect should be traced back to the
out-of-phase interference, contributed by the continuous dynamical phase
factors. To see this clearly, we first note that in Eq.~(\ref{SZZ1}) there
exists two kinds of phase factors, $\exp \left[ -\frac{i}{\hbar }%
(E_{N,+}-E_{L,+})t\right] $ and $\exp \left[ -\frac{i}{\hbar }%
(E_{N,+}-E_{L,-})t\right] $, of which the latter involves essential
information of the external fields, whereas the former does not. In fact,
for general values of $\mathcal{E}_{0}$ and $B_{0}$, the oscillating
frequencies of the latter, $(E_{N,+}-E_{L,-})/\hbar $, are nonzero and
usually significantly large. Thus, both the first and second terms in Eq.~(%
\ref{SZZ1}) vanish after long-time averaging, due to the out-of-phase
interference. However, when $\mathcal{E}_{0}$\ is tuned to some specific
values, such that
\begin{equation}
\mathcal{E}_{0}^{k}=\frac{1}{2\sqrt{\pi }\sigma _{t}\alpha e}\left[ k\hbar
\omega -\delta _{Z}\cos (\theta -\varphi )\right] ,  \label{PV}
\end{equation}%
with $k=L-N$, the system reaches its level (avoided) crossing point, i.e., $%
E_{N,+}-E_{L,-}=2\delta _{Z}\left\vert \eta \right\vert \sin (\theta
-\varphi )\approx 0$, which can be calculated from the second line of Eq.~(%
\ref{SPEC}). At this point, the out-of-phase interference is maximally
suppressed [the out-of-phase part vanishes if $E_{N,+}-E_{L,-}=0$], giving
rise to a considerable non-zero contribution to $Q$, and the
interference-resonant peaks thus emerge.

The tunability of the widths of the resonant peaks, which favors its
experimental observability, can be achieved by varying $T$. As shown in Fig.~2 by the red-dashed curve, a shorter averaging timespan, $T=5\pi
/\omega $, results in broader resonant peaks but does not change their peak
positions. We emphasize that the above analysis is general and independent
of the specific values of $\overline{A}_{N}$ and $\overline{B}_{N}$. That is
to say, different choices of the values of $c_{+}$ and $c_{-}$ of the
initial state, except for a special case $c_{+}\cdot c_{-}=0$, only lead to
different amplitudes of the resonant peaks rather than their specific
positions, which has also been confirmed by direct numerical simulations
(see Fig. \ref{FourInitial}). Our main results, therefore, are robust and
will not be affected.

From Eq.~(\ref{PV}), we find that the interval of any neighboring resonant
peaks,
\begin{equation}
\Delta \mathcal{E}_{k,k+1}=\mathcal{E}_{0}^{k+1}-\mathcal{E}_{0}^{k}=\frac{%
\hbar \omega }{2\sqrt{\pi }\sigma _{t}\alpha e},
\end{equation}%
remains a constant irrespective of the magnetic field, but depends crucially
on the SOC. However, the position of each individual resonant peak is
shifted by a small value via the Zeeman interaction, which is sensitive to
the direction of the magnetic field. The above two features of the proposed
interferometer provide a meaningful method to precisely measure relevant
experimental parameters. A typical example is the determination of the
Rashba and Dresselhaus SOC strengths, which is of critical importance in
current condensed-matter experiments\cite{CAL07,LG07,MSM10,MPW12}. To this
end, we first note that the total SOC strength is obtained explicitly
through $\alpha =\hbar \omega /(2\sqrt{\pi }\sigma _{t}e\Delta \mathcal{E}%
_{k,k+1})$ by measuring $\Delta \mathcal{E}_{k,k+1}$. Moreover, using the
relation between the resonant-peak position and the magnetic-field
direction, we can further determine the strengths of the Rashba and
Dresselhaus SOCs. In Fig.~\ref{Mdirection}, we numerically monitor the
response of the position of the second resonant-peak $\mathcal{E}_{0}^{1}$
on the magnetic-field direction $\theta $, by using $\delta _{Z}=0.06\hbar
\omega $. Bearing in mind the existence of a finite time to turn on the
magnetic field in a realistic situation, in our numerical simulation we have
intentionally replaced the Heaviside temporal shape of $B(t)$ by an
exponential-ramped timing, namely,
\begin{equation}
B(t)=B_{0}\left[ 1-\exp \left( -\frac{2.3(t-t_{0})}{\tau }\right) \right]
\Theta (t-t_{0}),  \label{RAB}
\end{equation}%
with $\tau $ being the timespan to turn on the magnetic field [the factor $%
2.3$ in Eq.~(\ref{RAB}) ensures that $B(t_{0}+\tau )/B_{0}=0.9$]. Obviously,
in the limit $\tau \rightarrow 0$, we recover the condition $%
B(t)=B_{0}\Theta (t-t_{0})$. As shown in Fig.~\ref{Mdirection}, for $\tau
=4.3/\omega $, the direct numerical simulations agree well with the
analytical expression in Eq.~(\ref{PV}), implying that the Heaviside
function $\Theta (t-t_{0})$ is a good approximation\ of a realistic
situation. Being similar to the previous results\cite{LRYJ13}, the position
of the resonant peak reaches its extreme value at $\theta _{m}=n\pi +\varphi
$ ($n=0,$ $\pm 1,$ $\pm 2,\cdots $), and $\varphi $ is thus determined from
the obtained $\theta _{m}$. Having obtained the auxiliary parameters $\alpha
$ and $\varphi $, the Rashba and Dresselhaus SOC strengths are expressed
directly as $\alpha _{R}=\alpha \tan \varphi $ and $\alpha _{D}=\alpha \cot
\varphi $, respectively. Obviously, a similar method can also be employed to
determine the Land\'{e} factor, since Eq.~(\ref{PV}) also contains this
basic information.

Note that in the case of $\theta =\theta _{m}$, the Hamiltonian (\ref{HTN})
is exactly solvable due to the cancelation of the redefined spin operator $%
\Sigma _{y}$. Further calculations accordingly prove the validity of Eqs.~(%
\ref{SPEC}) and (\ref{PV}), without any limitation on $B_{0}$. In this case,
using Eq.~(\ref{PV}), it seems that similar magnetic-field-driven resonant
peaks would arise at $B_{0}^{k}=$ $k\hbar \omega /(\sqrt{\pi }\sigma
_{t}\alpha eg_{e}\mu _{B})$, without introducing the pulsed electric field.
However, this straightforward derivation is not valid. Unlike the pulsed
electric field, which serves as a beam splitter in the proposed
interferometer, the magnetic field does not shift the original orbit (see
the expression of $\left\vert \psi _{n,\sigma }\right\rangle $), and thus $%
\overline{A}_{N}=\overline{B}_{N}=0$ for any $N\neq 0$ in Eq.~(\ref{SZZ1}).
This implies that all the dynamical phase factors in Eq.~(\ref{SZZ1}) cancel
out, and the interference-resonant peaks vanish. Therefore, it can be seen
that the pulsed electric field plays a unique role in inducing transitions
between the external orbital states, which is the key to switching on the
interferometric process.

Based on current experimental conditions of nanowire QDs, we now estimate
various relevant parameters to show the experimental feasibility of our
proposal. Consider the material parameters of GaAs\cite{Walser}, namely, $%
g_{e}=-0.44$, $\alpha =1.83\times 10^{-11}$ eV$\cdot $cm/$\hbar $, $%
m=0.067m_{0}$, where $m_{0}$ is the electron mass, and assume a weak trap
potential $\hbar \omega =9.1$ $\mu $eV, which can be controlled by gate
voltages\cite{SV12}. Therefore, the width of the electric-field pulse and
the timespan to turn on the magnetic field is estimated respectively as $%
\sigma _{t}=0.05/\omega \approx 4$ ps and $\tau =4.3/\omega \approx 300$ ps,
which is experimentally reasonable in view of the fact that ultrafast field
pulses about the order of picosecond have been reported\cite{ZMJ08}.
Accordingly, a viable averaging timespan can be chosen as $T=10\pi /\omega
=2.2$ ns, which is shorter than the spin dephasing time $T_{2}^{\ast }$ in
GaAs QD, which typically is $\sim $ $10$ ns\cite{HSI11}. This confirms the
observation of the predicted interferometric resonance within the spin
coherence time. In terms of the above parameters, the interval of the
resonant peaks plotted in Fig.~2 is also given by $\Delta
\mathcal{E}_{k,k+1}=\hbar \omega /(2\sqrt{\pi }\sigma _{t}\alpha e)=22.6$
V/cm. Since the Land\'{e} factor of GaAs is very small ($g_{e}=-0.44$), a
weak Zeeman energy, $\delta _{Z}=0.06\hbar \omega $, still supports a
considerably-strong magnetic field $B=42$ mT, which in turn shifts the
resonant peaks by, as large as, $\delta \mathcal{E}=\delta _{Z}/(2\sqrt{\pi }%
\sigma _{t}\alpha e)=1.35$ V/cm. Moreover, considering the fact that the
electric field in current experiments can easily reach $\sim $ $10^{5}$
V/cm, our interferometric method can thus be applied to precisely measure
materials with even much weaker SOC strengths. The relevant parameters of
some other semiconductor materials, namely GaAs, InSb, InAs, ZnO, and GaN,
can be found in Table \ref{table1}.\newline

\section*{{\protect\LARGE \textbf{Discussion \ }}}

Taking into account the fact that the anharmonicity is unavoidable in an
actual experiment, we add a higher order anharmonic factor, $\beta x^{4}$,
in the Hamiltonian (\ref{HTN}), i.e.,%
\begin{eqnarray}
H &=&\frac{p^{2}}{2m}+\frac{1}{2}m\omega ^{2}x^{2}+\beta x^{4}+\alpha \Sigma
_{x}p-\frac{e}{m}A(t)p  \notag \\
&&+\frac{1}{2}g_{e}\mu _{B}B(t)\left[ \Sigma _{x}\cos (\theta -\varphi
)+\Sigma _{y}\sin (\theta -\varphi )\right] ,  \label{HTD}
\end{eqnarray}%
to numerically analyze its impact on the interference peaks. To be clarity,
a dimensionless parameter, $\lambda =2\beta \hbar /(m^{2}\omega ^{3})$, is
introduced. Figure \ref{Anharmonic} shows the interference patterns with
respect to different $\lambda $. It can be seen that (i) the first few peaks
are stable if $\lambda $ is relatively small and with the increasing of $%
\lambda $, the peaks tend to deviate from their standard values of the
harmonic case; (ii) the higher resonant peaks are more sensitive to the
anharmonic perturbation than the lower ones. These behaviors can be
qualitatively explained as follows. The effect of the anharmonicity can be
neglected only if the quartic term $\left\langle \beta x^{4}\right\rangle $
is effectively small enough than $\left\langle m\omega
^{2}x^{2}/2\right\rangle $ ($\left\langle ...\right\rangle $ means
expectation value). With the increasing of $\mathcal{E}_{0}$, the wave
function is more likely to be excited to the higher orbit states (see the
expressions of $\overline{A}_{N}$ and $\overline{B}_{N}$) and therefore
becomes more extended, which consequently highlights the impact of the
quartic term $\left\langle \beta x^{4}\right\rangle $. On the other hand,
when the orders of $\lambda $ is no more than $10^{-3}$, the first two
resonant peaks are definitely stable, reflecting their robustness under the
anharmonicity. As we elucidated in the previous section, the knowledge of
the first two peaks is sufficient to extract information of the considered
SOC electron.

Note that various techniques have been employed to determine the Rashba and
Dresselhaus SOC strengths\cite{SV12,CAL07,LG07,MSM10,MPW12}. We emphasize
here, however, that our proposal is much different from previous schemes in
principle. Firstly, unlike previous works, in which the spin precession
responding to external magnetic fields is mainly investigated\cite%
{LG07,MSM10,MPW12}, the physics we exploited in this report essentially
originates from the interference of the dynamical phase factors of different
orbital states. The electric field here serves as a basic building block,
say, the beam splitter, of a multiple-polarization-interferometer. Secondly,
the existing schemes mostly extract information from the instantaneous spin
evolution, while the novel spin dynamics prediceted in this work reflects in
its long-time mean value.

Finally, we emphasize that the proposed idea of the SOC-induced
multiple-polarization quantum interferometer is general and the 1D nanowire
QD just offers a platform to demonstrate the relevant physics. Actually, the
model Hamiltonian (\ref{HT}) should, by no means, be limited to only a
single specific system. Resent advances in ultracold atoms, with artificial
gauge fields, make it another alternative candidate to exhibit the same
physics. For example, the harmonic-trapped two-component Bose-Einstein
Condensation (BEC), with synthetic 1D SOC, can be well simulated by the free
Hamiltonian (\ref{TH0})\cite{ISpielman,JD11}, and furthermore, a rapid shake
of the harmonic trapping potential of the BEC ideally corresponds to the
pulsed electric field $\mathcal{E}(t)$ employed in the nanowire QD. Thus,
following similar procedures we discussed above, a BEC-based interferometric
resonance, with respect to the strength of the shake, can also be expected.

In summary, we have theoretically formulated an electron-orbital
interferometry, by focusing on a 1D nanowire QD with SOC. By properly
adjusting the external fields' timing, different spin-orbit states are shown
to interfere with each other, generating intriguing interference-resonant
patterns. We have also shown that this interferometry has potential
applications in precise measuring relevant experimental parameters, such as
the Rashba and Dresselhaus SOC strengths, as well as the Land\'{e} factor.%
\newline

\section*{{\protect\LARGE \textbf{Methods}}}

\subsection{Derivation of the perturbed eigenstates in the
degenerate/quasi-degenerate case.}

By varying the parameters in Eq.~(\ref{EN}), it is possible to achieve a
regime where the $n$th and ($n+k$)th orbits are degenerate/quasi-degenerate,
say $\left\vert \epsilon _{n,+}-\epsilon _{n+k,-}\right\vert \ll \hbar
\omega $; see Fig.~\ref{Energy}. In such case, we must recombine the
unperturbed eigenstates $\left\vert \psi _{n}^{\sigma }\right\rangle
\left\vert \sigma \right\rangle $ to obtain proper zeroth-order eigenstates.

We assume that the $n$th perturbed eigenstate can be expressed as
\begin{equation}
\left\vert \psi _{n}^{(0)}\right\rangle =a\left\vert \psi
_{n}^{+}\right\rangle \left\vert +\right\rangle +b\left\vert \psi
_{n+k}^{-}\right\rangle \left\vert -\right\rangle .  \label{NPE}
\end{equation}%
Substituting the assumed eigenstate in Eq.~(\ref{NPE}) into the Schr\"{o}%
dinger equation $(H_{\text{S}}+H_{\text{Z}})\left\vert \psi
_{n}^{(0)}\right\rangle =E_{n}\left\vert \psi _{n}^{(0)}\right\rangle $ and
making use of $H_{\text{S}}\left\vert \psi _{n}^{\sigma }\right\rangle
\left\vert \sigma \right\rangle =$ $\epsilon _{n,\sigma }\left\vert \psi
_{n}^{\sigma }\right\rangle \left\vert \sigma \right\rangle $, we obtain the
following two equations:
\begin{equation}
\left[ \epsilon _{n,+}+\delta _{Z}\cos (\theta -\varphi )-E_{n}\right]
a+\eta ^{\ast }\delta _{Z}\sin (\theta -\varphi )b=0,  \label{Eq1}
\end{equation}%
\begin{equation}
\left[ \epsilon _{n+k,-}-\delta _{Z}\cos (\theta -\varphi )-E_{n}\right]
b+\eta \delta _{Z}\sin (\theta -\varphi )a=0.  \label{Eq2}
\end{equation}

The appearance of nonzero solutions in Eqs.~(\ref{Eq1}) and (\ref{Eq2})
requires%
\begin{equation}
\left\vert
\begin{array}{c}
\epsilon _{n,+}+\delta _{Z}\cos (\theta -\varphi )-E_{n},\text{ \ }\eta
^{\ast }\delta _{Z}\sin (\theta -\varphi ) \\
\eta \delta _{Z}\sin (\theta -\varphi ),\text{ \ }\epsilon _{n+k,-}-\delta
_{Z}\cos (\theta -\varphi )-E_{n}%
\end{array}%
\right\vert =0,  \label{NS}
\end{equation}%
which results in
\begin{equation}
E_{n,\pm }=n\hbar \omega \pm \frac{1}{2}(k\hbar \omega -f),  \label{EZ}
\end{equation}%
where%
\begin{equation}
f=\left\{ 4\left\vert \eta \right\vert ^{2}\delta _{Z}^{2}\sin ^{2}(\theta
-\varphi )+\left[ k\hbar \omega +2e\alpha A_{0}+2\delta _{Z}\cos (\theta
-\varphi )\right] ^{2}\right\} ^{\frac{1}{2}}.  \label{F}
\end{equation}%
The corresponding eigenstates are given by%
\begin{equation}
\left\vert \psi _{n,\pm }^{(0)}\right\rangle =a_{n,\pm }\left\vert \psi
_{n}^{+}\right\rangle \left\vert +\right\rangle +b_{n,\pm }\left\vert \psi
_{n+k}^{-}\right\rangle \left\vert -\right\rangle ,  \label{ZEST}
\end{equation}%
where
\begin{equation}
a_{n,\pm }=\pm \frac{2\eta ^{\ast }\delta _{Z}\sin (\theta -\varphi )}{\sqrt{%
4\left\vert \eta \right\vert ^{2}\delta _{Z}^{2}\sin ^{2}(\theta -\varphi )+%
\left[ f\mp 2\delta _{Z}\cos (\theta -\varphi )\right] ^{2}}},  \label{AN}
\end{equation}%
\begin{equation}
b_{n,\pm }=\frac{f\mp 2\delta _{Z}\cos (\theta -\varphi )}{\sqrt{4\left\vert
\eta \right\vert ^{2}\delta _{Z}^{2}\sin ^{2}(\theta -\varphi )+\left[ f\mp
2\delta _{Z}\cos (\theta -\varphi )\right] ^{2}}}.  \label{BN}
\end{equation}

Based on the rearranged zeroth-order eigenstates $\left\vert \psi _{n,\pm
}^{(0)}\right\rangle $ in Eq.~(\ref{ZEST}), the first-order eigenstates are
derived straightforwardly from perturbation theory. The results are given by%
\begin{equation}
\left\vert \psi _{n,\pm }\right\rangle =\left\vert \psi _{n,\pm
}^{(0)}\right\rangle +\overset{\infty }{\sum\limits_{l\neq n,n+k}}F_{1}^{\pm
}(l,n)\left\vert \psi _{l}^{+}\right\rangle \left\vert +\right\rangle +%
\overset{\infty }{\sum\limits_{l\neq n,n+k}}F_{2}^{\pm }(l,n)\left\vert \psi
_{l+k}^{-}\right\rangle \left\vert -\right\rangle ,  \label{RET}
\end{equation}%
where
\begin{eqnarray}
F_{1}^{\pm }(l,n) &=&\frac{a_{l,+}}{\epsilon _{n,\pm }-\epsilon _{l,+}}%
(a_{l,+}b_{n,\pm }\left\langle \psi _{l}^{+}\right\vert \left. \psi
_{n+k}^{-}\right\rangle +b_{l,+}a_{n,\pm }\left\langle \psi
_{l+k}^{-}\right\vert \left. \psi _{n}^{+}\right\rangle )+  \notag \\
&&\frac{a_{l,-}}{\epsilon _{n,\pm }-\epsilon _{l,-}}(a_{l,-}b_{n,\pm
}\left\langle \psi _{l}^{+}\right\vert \left. \psi _{n+k}^{-}\right\rangle
+b_{l,-}a_{n,\pm }\left\langle \psi _{l+k}^{-}\right\vert \left. \psi
_{n}^{+}\right\rangle ),  \label{F1}
\end{eqnarray}%
and%
\begin{eqnarray}
F_{2}^{\pm }(l,n) &=&\frac{b_{l,+}}{\epsilon _{n,\pm }-\epsilon _{l,+}}%
(a_{l,+}b_{n,\pm }\left\langle \psi _{l}^{+}\right\vert \left. \psi
_{n+k}^{-}\right\rangle +b_{l,+}a_{n,\pm }\left\langle \psi
_{l+k}^{-}\right\vert \left. \psi _{n}^{+}\right\rangle )+  \notag \\
&&\frac{b_{l,-}}{\epsilon _{n,\pm }-\epsilon _{l,-}}(a_{l,-}b_{n,\pm
}\left\langle \psi _{l}^{+}\right\vert \left. \psi _{n+k}^{-}\right\rangle
+b_{l,-}a_{n,\pm }\left\langle \psi _{l+k}^{-}\right\vert \left. \psi
_{n}^{+}\right\rangle ).  \label{F22}
\end{eqnarray}

\subsection{Complete expression of Eq.~(\protect\ref{SZZ1}) in the Results
Section.}

The detailed expressions of $M_{\alpha \beta }(N,L)$ ($\alpha ,\beta =\pm $)
in Eq.~(\ref{SZZ1}) should be divided into the following two cases: (i)
non-degenerate case, where each orbital energy $\epsilon _{n,\sigma }$ is
well separated from others, and (ii) degenerate/quasi-degenerate case, where
$\left\vert \epsilon _{n,+}-\epsilon _{n+k,-}\right\vert \ll \hbar \omega $.
Specially, in the non-degenerate case we have
\begin{equation}
M_{++}(N,L)=-i\overline{A}_{N}\overline{A}_{L}^{\ast }\delta _{Z}\sin
(\theta -\varphi )\overset{\infty }{\sum\limits_{n\neq N}}\frac{\left\langle
\psi _{n}^{-}\right\vert \left. \psi _{N}^{+}\right\rangle \left\langle \psi
_{L}^{+}\right\vert \left. \psi _{n}^{-}\right\rangle }{(N-n)\hbar \omega
-2m\alpha A_{0}},
\end{equation}%
\begin{equation}
M_{--}(N,L)=-i\overline{B}_{N}\overline{B}_{N}^{\ast }\delta _{Z}\sin
(\theta -\varphi )\overset{\infty }{\sum\limits_{n\neq L}}\frac{\left\langle
\psi _{L}^{-}\right\vert \left. \psi _{n}^{+}\right\rangle \left\langle \psi
_{n}^{+}\right\vert \left. \psi _{N}^{-}\right\rangle }{(L-n)\hbar \omega
+2m\alpha A_{0}},
\end{equation}%
\begin{eqnarray}
M_{+-}(N,L) &=&-i\overline{A}_{N}\overline{B}_{L}^{\ast }\delta _{Z}^{2}\sin
^{2}(\theta -\varphi )\overset{\infty }{\sum\limits_{n\neq N}}\overset{%
\infty }{\sum\limits_{\tilde{n}\neq L}}\frac{\left\langle \psi
_{n}^{-}\right\vert \left. \psi _{N}^{+}\right\rangle }{(N-n)\hbar \omega
-2m\alpha A_{0}} \\
&&\times \frac{\left\langle \psi _{L}^{-}\right\vert \left. \psi _{\tilde{n}%
}^{+}\right\rangle \left\langle \psi _{\tilde{n}}^{+}\right\vert \left. \psi
_{n}^{-}\right\rangle }{(L-\tilde{n})\hbar \omega +2m\alpha A_{0}},  \notag
\end{eqnarray}%
\begin{equation}
M_{-+}(N,L)=-i\overline{A}_{L}^{\ast }\overline{B}_{N}\left\langle \psi
_{L}^{+}\right\vert \left. \psi _{N}^{-}\right\rangle ,
\end{equation}%
and in the degenerate/quasi-degenerate case we obtain
\begin{eqnarray}
M_{+-}(N,L) &=&A_{N}B_{L}^{\ast }(-ia_{L,-}^{\ast }b_{N,+}\left\langle \psi
_{L}^{+}\right\vert \left. \psi _{N+k}^{-}\right\rangle -i\overset{\infty }{%
\sum\limits_{n\neq N}}a_{L,-}^{\ast }F_{2}^{+}(n,N)\left\langle \psi
_{L}^{+}\right\vert \left. \psi _{n+k}^{-}\right\rangle  \notag \\
&&-i\overset{\infty }{\sum\limits_{\tilde{n}\neq L}}b_{N,+}F_{1}^{-}(\tilde{n%
},L)\left\langle \psi _{\tilde{n}}^{+}\right\vert \left. \psi
_{N+k}^{-}\right\rangle +\overset{\infty }{\sum\limits_{\tilde{n}\neq L}}%
\overset{\infty }{\sum\limits_{n\neq N}}F_{1}^{-\ast }(\tilde{n}%
,L)F_{2}^{+}(n,N)\left\langle \psi _{\tilde{n}}^{+}\right\vert \left. \psi
_{n+k}^{-}\right\rangle  \notag \\
&&+i\overset{\infty }{\sum\limits_{\tilde{n}\neq L}}a_{N,+}^{\ast }F_{2}^{-}(%
\tilde{n},L)\left\langle \psi _{\tilde{n}+k}^{-}\right\vert \left. \psi
_{N}^{+}\right\rangle -\overset{\infty }{\sum\limits_{\tilde{n}\neq L}}%
\overset{\infty }{\sum\limits_{n\neq N}}F_{2}^{-\ast }(\tilde{n}%
,L)F_{1}^{+}(n,N)\left\langle \psi _{\tilde{n}+k}^{-}\right\vert \left. \psi
_{n}^{+}\right\rangle )  \notag \\
&&+ib_{L,-}^{\ast }a_{N,+}\left\langle \psi _{L+k}^{-}\right\vert \left.
\psi _{N}^{+}\right\rangle +i\overset{\infty }{\sum\limits_{n\neq N}}%
b_{L,-}^{\ast }F_{1}^{+}(n,N)\left\langle \psi _{L+k}^{-}\right\vert \left.
\psi _{n}^{+}\right\rangle ,  \label{M1}
\end{eqnarray}%
\begin{eqnarray}
M_{--}(N,L) &=&B_{L}^{\ast }B_{N}(-ia_{L,-}^{\ast }b_{N,-}\left\langle \psi
_{L}^{+}\right\vert \left. \psi _{N+k}^{-}\right\rangle -i\overset{\infty }{%
\sum\limits_{n\neq N}}a_{L,-}^{\ast }F_{2}^{-}(n,N)\left\langle \psi
_{L}^{+}\right\vert \left. \psi _{n+k}^{-}\right\rangle  \notag \\
&&-i\overset{\infty }{\sum\limits_{\tilde{n}\neq L}}b_{N,+}F_{1}^{-\ast }(%
\tilde{n},L)\left\langle \psi _{\tilde{n}}^{+}\right\vert \left. \psi
_{N+k}^{-}\right\rangle -i\overset{\infty }{\sum\limits_{\tilde{n}\neq L}}%
\overset{\infty }{\sum\limits_{n\neq N}}F_{1}^{-\ast }(\tilde{n}%
,L)F_{2}^{-}(n,N)\left\langle \psi _{\tilde{n}}^{+}\right\vert \left. \psi
_{n+k}^{-}\right\rangle  \notag \\
&&+i\overset{\infty }{\sum\limits_{\tilde{n}\neq L}}a_{N,-}^{\ast }F_{2}^{-}(%
\tilde{n},L)\left\langle \psi _{\tilde{n}+k}^{-}\right\vert \left. \psi
_{N}^{+}\right\rangle +i\overset{\infty }{\sum\limits_{\tilde{n}\neq L}}%
\overset{\infty }{\sum\limits_{n\neq N}}F_{2}^{-\ast }(\tilde{n}%
,L)F_{1}^{-}(n,N)\left\langle \psi _{\tilde{n}+k}^{-}\right\vert \left. \psi
_{n}^{+}\right\rangle )  \notag \\
&&+ib_{L,-}^{\ast }a_{N,-}\left\langle \psi _{L+k}^{-}\right\vert \left.
\psi _{N}^{+}\right\rangle +i\overset{\infty }{\sum\limits_{n\neq N}}%
b_{L,-}^{\ast }F_{1}^{-}(n,N)\left\langle \psi _{L+k}^{-}\right\vert \left.
\psi _{n}^{+}\right\rangle ,  \label{M2}
\end{eqnarray}%
\begin{eqnarray}
M_{-+}(N,L) &=&A_{L}^{\ast }B_{N}(-ia_{L,+}^{\ast }b_{N,-}\left\langle \psi
_{L}^{+}\right\vert \left. \psi _{N+k}^{-}\right\rangle -i\overset{\infty }{%
\sum\limits_{n\neq N}}a_{L,+}^{\ast }F_{2}^{-}(n,N)\left\langle \psi
_{L}^{+}\right\vert \left. \psi _{n+k}^{-}\right\rangle  \notag  \label{M3}
\\
&&-i\overset{\infty }{\sum\limits_{\tilde{n}\neq L}}b_{N,-}F_{1}^{+\ast }(%
\tilde{n},L)\left\langle \psi _{\tilde{n}}^{+}\right\vert \left. \psi
_{N+k}^{-}\right\rangle +\overset{\infty }{\sum\limits_{\tilde{n}\neq L}}%
\overset{\infty }{\sum\limits_{n\neq N}}F_{1}^{+\ast }(\tilde{n}%
,L)F_{2}^{-}(n,N)\left\langle \psi _{\tilde{n}}^{+}\right\vert \left. \psi
_{n+k}^{-}\right\rangle  \notag \\
&&+i\overset{\infty }{\sum\limits_{\tilde{n}\neq L}}a_{N,-}F_{2}^{+\ast }(%
\tilde{n},L)\left\langle \psi _{\tilde{n}+k}^{-}\right\vert \left. \psi
_{N}^{+}\right\rangle -\overset{\infty }{\sum\limits_{\tilde{n}\neq L}}%
\overset{\infty }{\sum\limits_{n\neq N}}F_{2}^{+\ast }(\tilde{n}%
,L)F_{1}^{-}(n,N)\left\langle \psi _{\tilde{n}+k}^{-}\right\vert \left. \psi
_{n}^{+}\right\rangle )  \notag \\
&&+ib_{L,+}^{\ast }a_{N,-}\left\langle \psi _{L+k}^{-}\right\vert \left.
\psi _{N}^{+}\right\rangle +i\overset{\infty }{\sum\limits_{n\neq N}}%
b_{L,+}^{\ast }F_{1}^{+}(n,N)\left\langle \psi _{L+k}^{-}\right\vert \left.
\psi _{n}^{+}\right\rangle ,  \label{M33}
\end{eqnarray}%
\begin{eqnarray}
M_{++}(N,L) &=&A_{L}^{\ast }A_{N}(-ia_{L,+}^{\ast }b_{N,+}\left\langle \psi
_{L}^{+}\right\vert \left. \psi _{N+k}^{-}\right\rangle -i\overset{\infty }{%
\sum\limits_{n\neq N}}a_{L,+}^{\ast }F_{2}^{+}(n,N)\left\langle \psi
_{L}^{+}\right\vert \left. \psi _{n+k}^{-}\right\rangle  \notag \\
&&-i\overset{\infty }{\sum\limits_{\tilde{n}\neq L}}b_{N,+}F_{1}^{+\ast }(%
\tilde{n},L)\left\langle \psi _{\tilde{n}}^{+}\right\vert \left. \psi
_{N+k}^{-}\right\rangle +\overset{\infty }{\sum\limits_{\tilde{n}\neq L}}%
\overset{\infty }{\sum\limits_{n\neq N}}F_{1}^{+\ast }(\tilde{n}%
,L)F_{2}^{+}(n,N)\left\langle \psi _{\tilde{n}}^{+}\right\vert \left. \psi
_{n+k}^{-}\right\rangle  \notag \\
&&+i\overset{\infty }{\sum\limits_{\tilde{n}\neq L}}a_{N,+}F_{2}^{+\ast }(%
\tilde{n},L)\left\langle \psi _{\tilde{n}+k}^{-}\right\vert \left. \psi
_{N}^{+}\right\rangle +\overset{\infty }{\sum\limits_{\tilde{n}\neq L}}%
\overset{\infty }{\sum\limits_{n\neq N}}F_{2}^{+\ast }(\tilde{n}%
,L)F_{1}^{+}(n,N)\left\langle \psi _{\tilde{n}+k}^{-}\right\vert \left. \psi
_{n}^{+}\right\rangle )  \notag \\
&&+ib_{L,+}^{\ast }a_{N,+}\left\langle \psi _{L+k}^{-}\right\vert \left.
\psi _{N}^{+}\right\rangle +i\overset{\infty }{\sum\limits_{n\neq N}}%
b_{L,+}^{\ast }F_{1}^{+}(n,N)\left\langle \psi _{L+k}^{-}\right\vert \left.
\psi _{n}^{+}\right\rangle .  \label{M4}
\end{eqnarray}%
\newline

\section*{{\protect\LARGE \textbf{Acknowledgments}}}

We thank Dr.~Rui Li for numerous insightful discussions. This work is
supported in part by the NSFC under Grants No.~61275211, No.~11422433,
No.~11434007, No.~61574087, and No.~11674200; the PCSIRT under Grant
No.~IRT13076; the NCET under Grant No.~13-0882; the FANEDD under Grant
No.~201316; OYTPSP; and SSCC. It was also partially supported by the RIKEN
iTHES Project, the MURI Center for Dynamic Magneto-Optics via the AFOSR
Award No.~FA9550-14-1-0040, and a Grant-in-Aid for Scientific Research (A).

\section*{{\protect\LARGE \textbf{Author Contributions}}}

J.F., Y.C., G.C., L.X., S.J. and F.N. conceived the idea, J.F., Y.C., L.X.
and G.C. performed the calculation, G.C., S.J. and F.N. wrote the
manuscript, G.C. supervised the whole research project.

\section*{{\protect\LARGE \textbf{Additional information}}}

\textbf{Competing financial interests:} The authors declare no competing
financial interests.

\newpage

\textbf{Figure 1: Schematic description of the proposed interferometer.} Top
panel: The initial spin-orbit state, staying in the zeroth orbit, is split
into different orbital states at time $t_{0}$. Each orbital state evolves
independently in its individual passage, accumulating a relative phase
shift. The interference pattern arises from a specific measurement, which
also acts as a beam recombiner. Bottom panel: Timing of the pulsed electric
field, which shifts the original orbit and plays the role of a beam splitter.

\textbf{Figure 2: The long-time averaged spin polarization $Q$ versus the
peak amplitude }$\mathcal{E}_{0}$\textbf{.} Here, the SOC strength, the
temporal width, and the initial state are given by $\alpha =0.08\sqrt{\hbar
\omega /m}$, $\sigma _{t}=0.05/\omega $, and $\left\vert \Psi
(0)\right\rangle =$ $\frac{1}{\sqrt{2}}(\left\vert \phi
_{0}^{+}\right\rangle \left\vert +\right\rangle +\left\vert \phi
_{0}^{-}\right\rangle \left\vert -\right\rangle )$, respectively. The
timespans for the red-dashed and blue-solid curves are chosen as $T=5\pi
/\omega $ and $40\pi /\omega $, respectively.

\textbf{Figure 3: The long-time averaged spin polarization $Q$ versus the
peak amplitude }$\mathcal{E}_{0}$\textbf{\ for four different initial states.%
} In these plots, $c_{+}/c_{-}$ is given by (a) $2/1$, (b) $3/1$, (c) $2/3$,
and (d) $3/4$, respectively. The timespan is chosen as $T=10\pi /\omega $.
Other parameters are the same as those in Fig.~2.

\textbf{Figure 4: The position of the second resonant-peak }$\mathcal{E}$%
\textbf{$_{0}^{1}$ versus the magnetic-field direction $\theta $.} Here, the
timespan and the timespan to turn on the magnetic field are given by $%
T=10\pi /\omega $ and $\tau =4.3/\omega $, respectively. The SOC strength $%
\alpha $, the temporal width $\sigma _{t}$, and the initial state $%
\left\vert \Psi (0)\right\rangle $ are the same as those in Fig.~2. The black open circles correspond to direct numerical
simulations, while the curves show our analytical results in Eq.~(\ref{PV}).

\textbf{Figure 5: The long-time averaged spin polarization $Q$ versus the
peak amplitude }$\mathcal{E}$\textbf{$_{0}$ with respect to different
anharmonic factors, }$\lambda =10^{-3}$ \textbf{(red dashed curve),} $%
\lambda =10^{-4}$ \textbf{(blue dotted-dashed curve) and} $\lambda =10^{-5}$
\textbf{(black solid curve).} The timespan is chosen as $T=10\pi /\omega $.
Other parameters are the same as those in Fig.~2.

\textbf{Figure 6: Schematic energy levels of the $n$th and ($n+k$)th orbits
in the nondegenerate case (NC) and degenerate/quasi-degenerate case (DC).}

\begin{figure}[t]
\centering\includegraphics[width = 6.3in]{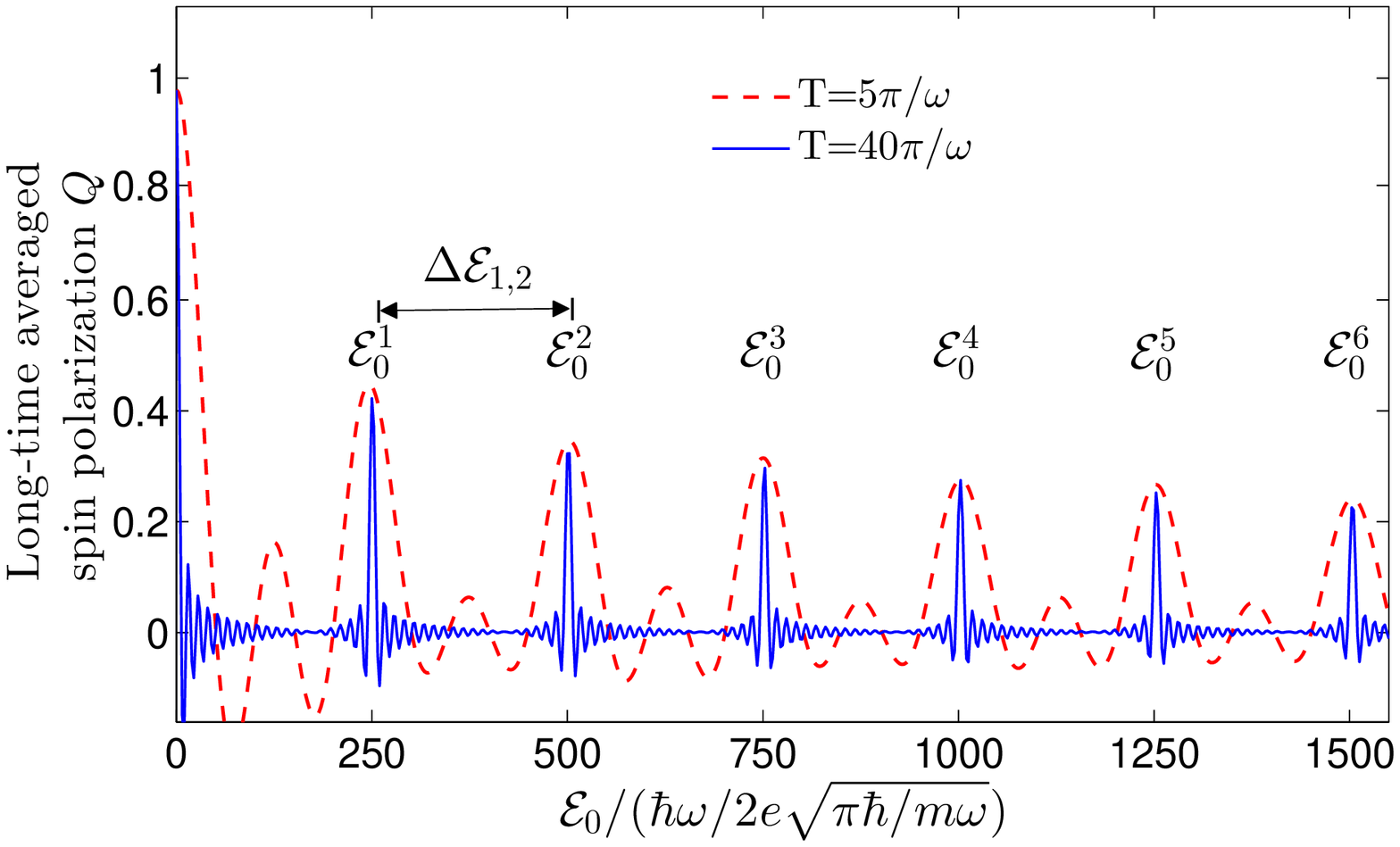}
\caption{\textbf{The long-time averaged spin polarization $Q$ versus the
peak amplitude $\mathcal{E}_{0}$.}}
\label{ResPeak}
\end{figure}

\begin{figure}[t]
\centering\includegraphics[width = 6.3in]{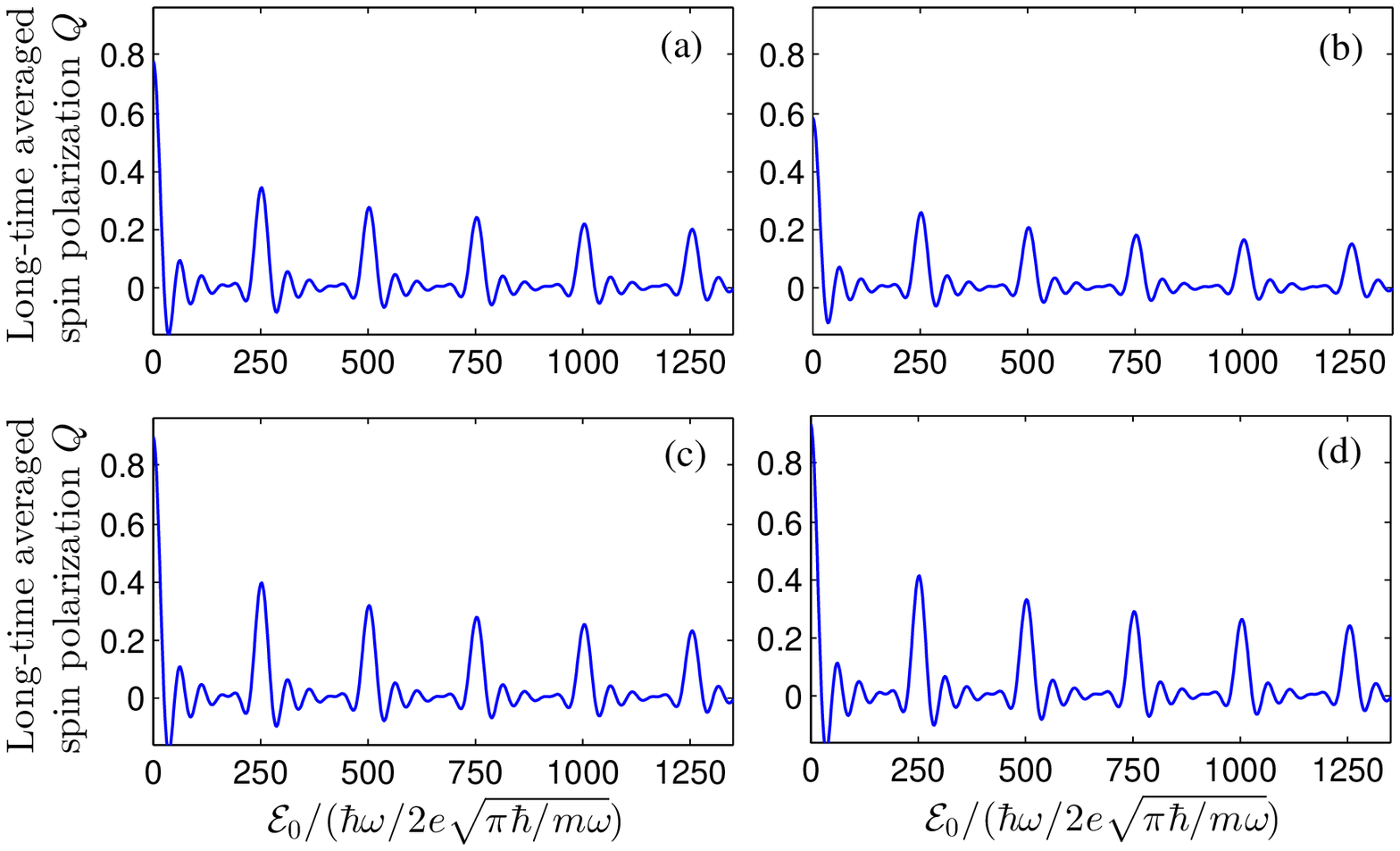}
\caption{\textbf{The long-time averaged spin polarization $Q$ versus the
peak amplitude $\mathcal{E}_{0}$ for four different initial states.}}
\label{FourInitial}
\end{figure}

\begin{figure}[t]
\centering\includegraphics[width = 6.3in]{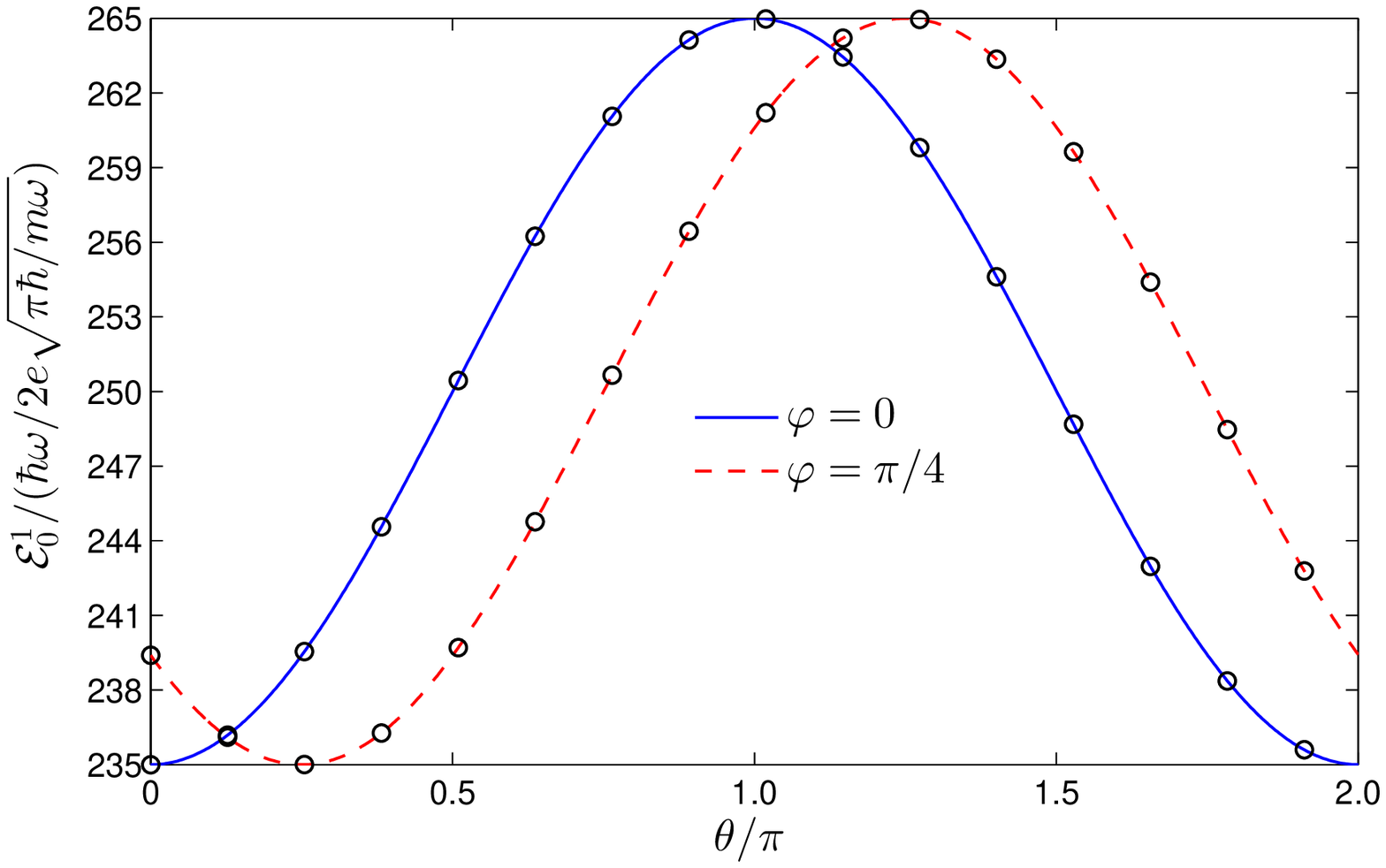}
\caption{\textbf{The position of the second resonant-peak $\mathcal{E}%
_{0}^{1}$ versus the magnetic-field direction $\protect\theta $.}}
\label{Mdirection}
\end{figure}

\begin{figure}[t]
\centering\includegraphics[width = 6.3in]{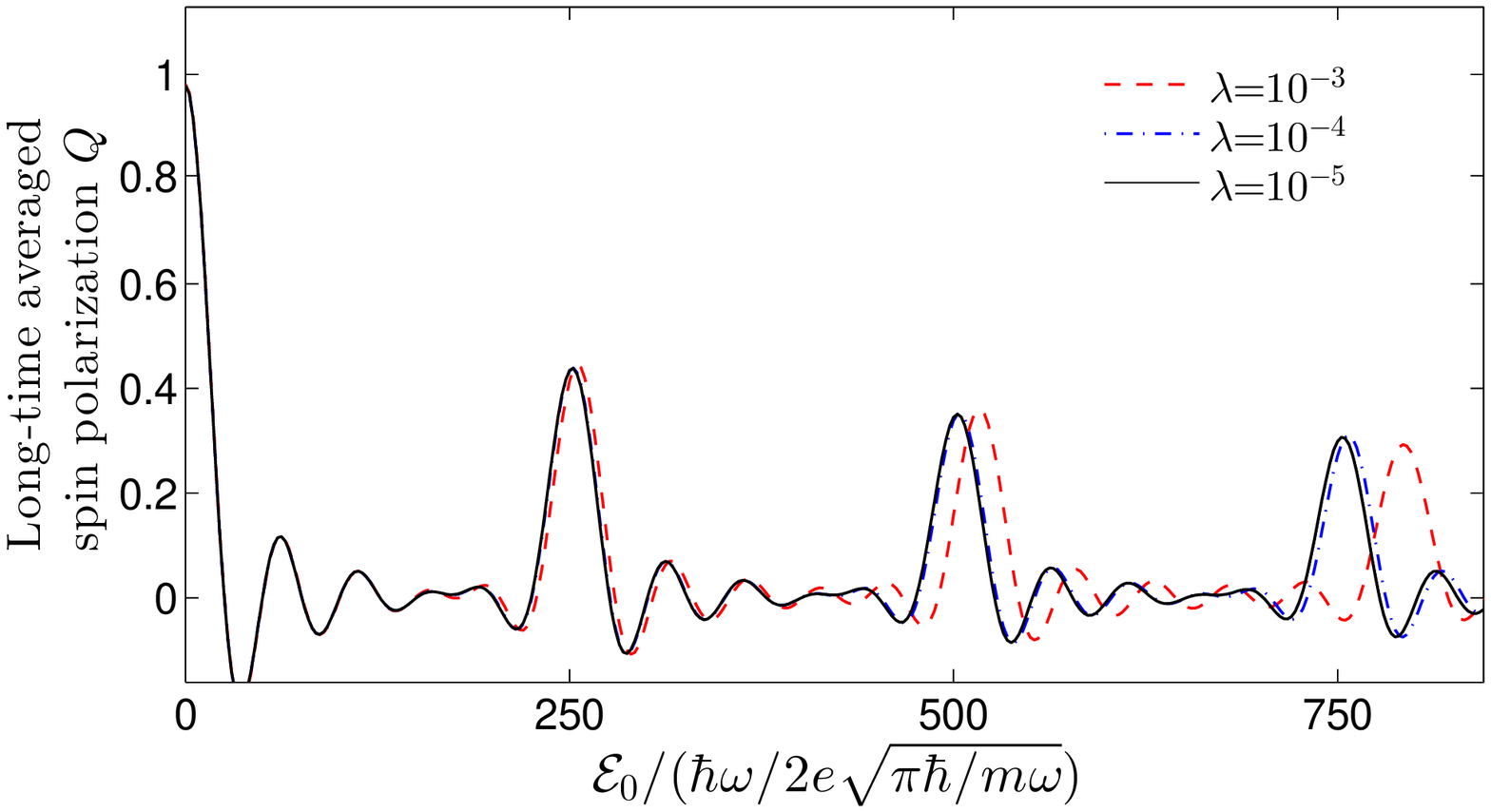}
\caption{\textbf{The long-time averaged spin polarization $Q$ versus the
peak amplitude $\mathcal{E}_{0}$ with respect to different anharmonic
factors, }$\protect\lambda =10^{-3}$ \textbf{(red dashed curve),} $\protect%
\lambda =10^{-4}$ \textbf{(blue dotted-dashed curve) and} $\protect\lambda %
=10^{-5}$ \textbf{(black solid curve).}}
\label{Anharmonic}
\end{figure}

\begin{figure}[t]
\centering\includegraphics[width = 6.3in]{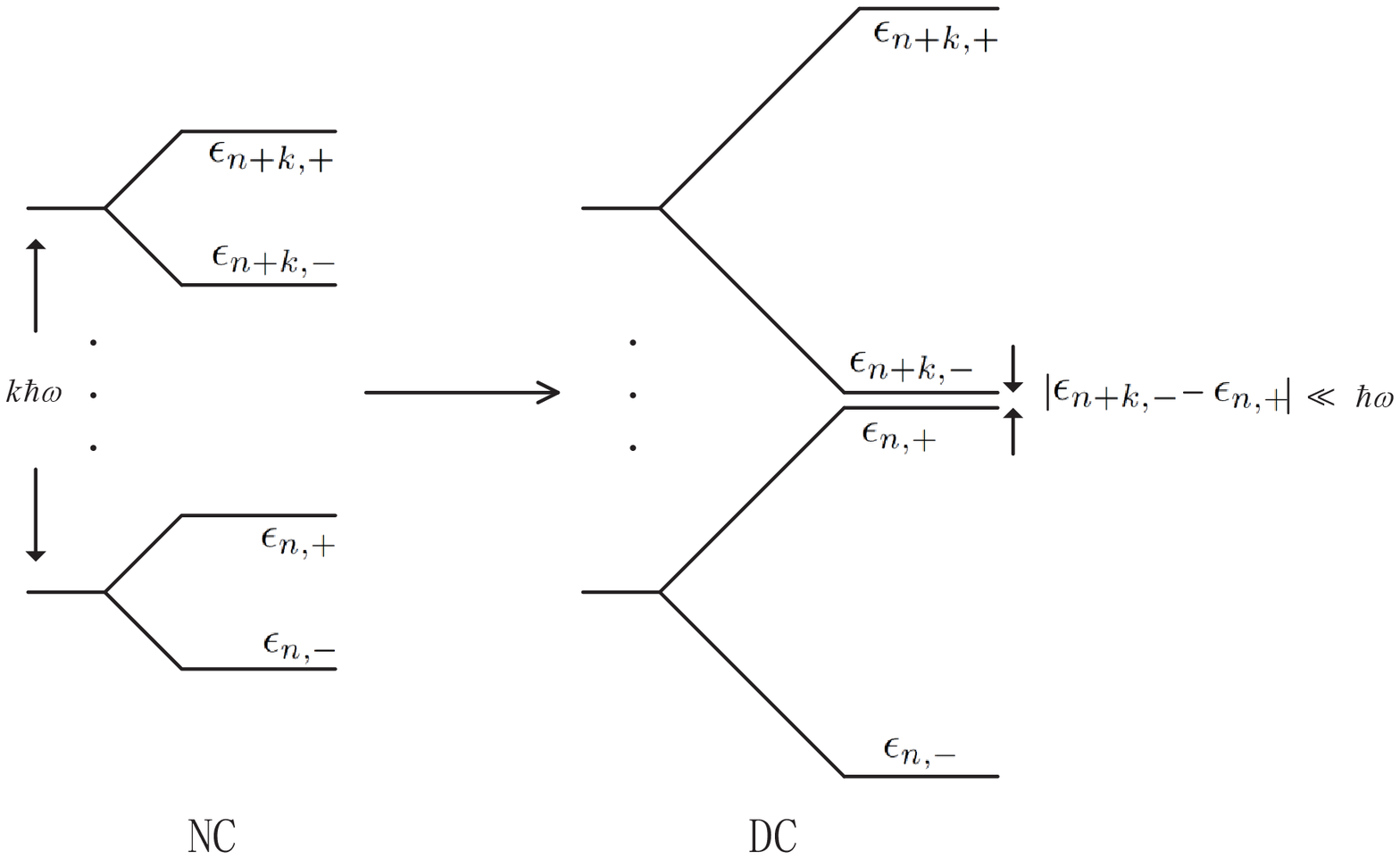}
\caption{\textbf{Schematic energy levels of the $n$th and ($n+k$)th orbits
in the nondegenerate case (NC) and degenerate/quasi-degenerate case (DC).}}
\label{Energy}
\end{figure}

\begin{table}[t]
\begin{tabular}{cccccc}
\hline\hline
Semiconductor & $\hbar \alpha _{R}$ (eV$\cdot $cm) & $\hbar \alpha _{D}$ (eV$%
\cdot $cm) & $\gamma $ (eV$\cdot $\AA $^{3}$) & $\hbar \alpha $ (eV$\cdot $%
cm) & $\Delta \mathcal{E}_{k,k+1}$ (V/cm) \\ \hline
GaAs & 0.68$\times $10$^{-11\text{ }\mbox{a}}$ & -1.7$\times $10$^{-11\text{
}}$ & -11$^{\mbox{a}}$ & 1.83$\times $10$^{-11}$ & 22.6 \\
$\,$InSb & 3$\times $10$^{-10\text{ }\mbox{b}}$ & 7.7$\times $10$^{-10\text{
}}$ & 490$^{\mbox{b}}$ & 8.3$\times $10$^{-10}$ & 0.5 \\
$\,$InAs & 5.71$\times $10$^{-9\text{ }\mbox{c}}$ & 9.0$\times $10$^{-10%
\text{ }}$ & 571.8$^{\mbox{c}}$ & 5.78$\times $10$^{-9}$ & 0.07 \\
$\,$ZnO & 1.1$\times $10$^{-11\text{ }\mbox{d}}$ & 5.2$\times $10$^{-13}$ &
0.33$^{\mbox{e}}$ & 1.1$\times $10$^{-11}$ & 37.6 \\
$\,$CaN & 9.0$\times $10$^{-11\text{ }\mbox{f}}$ & 5.0$\times $10$^{-13}$ &
0.32$^{\mbox{e}}$ & 9.0$\times $10$^{-11}$ & 4.6 \\ \hline\hline
\end{tabular}%
\newline
\centering$^{\mbox{a}}$Ref.~[50]; $^{\mbox{b}}$Ref.~[53]; $^{\mbox{c}}$%
Ref.~[54]; $^{\mbox{d}}$Ref.~[55]; $^{\mbox{e}}$Ref.~[56]; $^{\mbox{f}}$%
Ref.~[57];
\caption{\textbf{Some quantum dot parameters}.The Dresselhaus SOC strength
is estimated by $\hbar \protect\alpha _{D}\approx \protect\gamma (\protect%
\pi /z_{0})^{2}$, where $\protect\gamma $ is the material-specific constant
and $z_{0}$ is the quantum well vertical width. In these estimations, we
assume $z_{0}=25$ nm, $\hbar \protect\omega =9.1$ $\protect\mu $eV and $%
\protect\sigma _{t}=4$ Ps.}
\label{table1}
\end{table}

\end{document}